\documentstyle[12pt,aasms4]{article}

\lefthead{Stiavelli}
\righthead{Violent Relaxation and Black Holes}

\begin{document}

\title{Violent Relaxation Around a Massive Black Hole}

\author{M.~Stiavelli
\footnote{on assignment from the Space Science
Department of the European Space Agency}}

\affil{Space Telescope Institute, 3700 San Martin Drive, Baltimore, MD
21218, USA \\
and\\
Scuola Normale Superiore, Piazza dei Cavalieri, I56126, Pisa, Italy}

\begin{abstract}
I present galaxy models resulting from violent relaxation in the
presence of a pre-existing black hole. The models are computed by
maximizing the entropy of the stellar dynamical system. I show that
their properties are very similar to those of adiabatic growth models
for a suitable choice of parameters. This suggests that observations
of nuclear light profiles and kinematics alone may not be sufficient
to discriminate between scenarios where a black hole grows
adiabatically in the core of a galaxy, and scenarios where the black
hole formation preceeds galaxy formation.
\end{abstract}

{\it subject headings}: galaxies: elliptical and lenticular, cD ---
galaxies: nuclei --- galaxies: formation

\section{Introduction}

The search for massive black holes in the nuclei of galaxies began
with the pioneering work of Young and collaborators (1978, 1980), who
computed the effect of the adiabatic growth of a black hole on the
underlying stellar population of a galaxy, represented by an
isothermal sphere, and applied these models to M87. Further progress
in stellar dynamics uncovered a rich variety of possible stellar
equilibria (see, e.g., Binney and Mamon 1982), and urged for more
sophisticated studies to certify the validity of black hole
identifications based on simple models. The existence of black holes
in the cores of galaxies is now unambiguously established for the
nucleus of NGC 4258, where the rotation curve from water maser
emission has been measured down to the parsec scale (Miyoshi et
al. 1995). This has changed the prevalent attitude from that of {\it
proving} that black holes existed to that of {\it understanding} what
is the interplay between their properties and those of the nuclei
where they reside.

On the observational side, the refurbished Hubble Space Telescope has
provided data on the cores of galaxies with unprecedented angular
resolution (e.g., Harms et al. 1994; van der Marel et al. 1997a,b;
Kormendy et al.  1995, 1996; Macchetto et al. 1997). On the
theoretical side, models constructed with the Schwarzschild (1979)
technique have allowed one to confirm the generality of the
conclusions based on simpler models (van der Marel et al. 1998).  The
combination of photometry and spectroscopy with HST, and
state-of-the-art dynamical modeling, has revealed that adiabatic black
hole growth in isothermal cores may provide a fairly good
representation of the data (van der Marel et al. 1998; van der Marel
1998). The adiabatic growth scenario implies that black holes are
grown mostly from external gas rather than from disrupting stars.
This has wide-ranging consequences on the way black holes and active
galactic nuclei, and ultimately galaxies, form and evolve.

The question remains whether the adiabatic black hole growth is a {\it
unique} representation of the observed galactic nuclear properties,
i.e., whether a generally good fit by adiabatic models does provide a
decisive constraint about how galactic nuclei formed. To explore this
issue, one can investigate the physically opposite formation scenario
where the massive black hole pre-exists, and the galaxy forms around
it via a process of violent relaxation. Numerical studies have been
inconclusive for a variety of reasons, including, e.g., their small
number of particles (Dekel, Kowitt, and Shaham 1981), grid effects in
the neighbourhood of the central mass concentration (Udry 1993), or
because focussed mostly on the isophotal effects of BH growth (Norman,
May and van Albada 1985). In this letter I follow a different
approach. In Section 2 I present a brief summary of how the
statistical mechanics of violent relaxation can be used to derive the
core properties of a galaxy. In Section 3 I describe the
violent-relaxation models, which I compare to the adiabatic
models. Section 4 sums up.

\section{Statistical Mechanics of Violent Relaxation}

Violent relaxation allows a collisionless stellar dynamical system to
reach equibrium by means of strong fluctuations of the mean
gravitational potential.  In the absence of {\it degeneracy} (i.e., if
the phase space density of the system is everywhere well below the
maximum allowed by Liouville's theorem), the end-product of violent
relaxation is an isothermal distribution function (Lynden-Bell 1967,
hereafter LB67; Shu 1978). Given that such distribution function is
characterized by an unphysical infinite mass, suitable truncations
have been investigated.  In his seminal paper, LB67 already pointed
out the astrophysical problem of the {\it incompleteness} of violent
relaxation as the solution to the problem of the unphysical infinite
mass of the maximum entropy distribution function. Other authors since
then have investigated the incompleteness paradigm with a variety of
methods (e.g., Tremaine, H{\'e}non and Lynden-Bell 1986; Stiavelli \&
Bertin 1987; Stiavelli 1987; Madsen 1987, Shu 1987; Hjort and Madsen
1991;).

A second important aspect of the violent relaxation distribution
function, which was also pointed out by LB67 but has received less
attention, is that of {\it degeneracy}. Here I focus on this issue.
To do so, I first rederive the expression of the maximum entropy
distribution function by including also an external potential. In the
notation of Stiavelli \& Bertin (1987; see also Shu 1978) I define a
microstate as a set of occupation numbers in 6-dimensional microcells
small enough that at most one star will occupy each microcell. This is
readily ensured by selecting $g=1/p_{max}$ as the microcell size, with
$p_{max}$ the maximum initial value of the distribution function
before collapse, merging, or any other violent relaxation mechanism
which is provided by nature at galaxy formation. In a collisionless
formation scenario, Liouville's theorem guarantees that in its
subsequent evolution, the distribution function will never exceed this
value. One can now partition the phase space into coarser macrocells,
with each macrocell containing $\nu_a$ microcells.  The observable
state of the system will depend only on the occupation numbers ${n_a}$
of these macrocells, which we term a macrostate. In general $0 \leq
n_a \leq min({\nu_a, N})$, with $N$ the total number of stars in the
system. Let $M$ be the total number of macrocells.

The entropy of the system is $S=log W( {n_a} )$ where
\begin{equation}
W( {n_a} ) = \frac{N!}{n_1!...n_M!} \frac{\nu_1!...\nu_M!}
{(\nu_1-n_1)!...(\nu_M-n_M)!}
\end{equation}
is the number of microstates associated with a given macrostate. By
using the Stirling approximation one finds:
\begin{equation}
\label{entropyeq}
S = -\sum_{a=1}^{M} n_a \log{n_a} - \sum_{a=1}^{M} (\nu_a-n_a) 
\log{(\nu_a-n_a)} + const.
\end{equation}

The entropy is maximized under the constraints of the conservation of
the total energy and number of particles, namely:
\begin{equation}
\label{energyeq}
E_{tot} = \sum_{a=1}^M m n_a (\frac{v_a^2}{2}+\frac{1}{2}\Phi_a^{(int)}
+\Phi_a^{(ext)})
\end{equation}
and
\begin{equation}
\label{numbereq}
N = \sum_{a=1}^M n_a
\end{equation}
where in the expression for the energy I have separated the
gravitational potential in an internal component due to self-gravity,
$\Phi_a^{(int)}$, and an external component, $\Phi_a^{(ext)}$, and
have indicated with $m$ and $v_a$ the particle masses and velocities,
respectively. The resulting maximum entropy distribution function is
given by:
\begin{equation}
\label{distribfuneq}
n_a = \frac{\nu_a}{1+\exp{(\beta E_a -\mu)}}
\end{equation}
with $E_a = \frac{1}{2}{m v_a^2}+ m \Phi_a^{(int)}+m \Phi_a^{(ext)}$
the energy per particle, and $\beta$ and $\mu$ the Lagrange
multipliers for the energy and number of particles, respectively. The
ratio $\mu/\beta$ is the chemical potential of the system. For a
Fermi-Dirac type distribution function such as that of
Eq. \ref{distribfuneq}, the chemical potential can be either positive
or negative. Its value represents the so-called Fermi energy, i.e.
the upper limit to the individual particle energy of completely
degenerate systems.

Whenever $1 << \exp{(\beta E_a -\mu)}$, one has $n_a << \nu_a$,
i.e.  most of the microcells are not occupied, and the system is
non-degenerate. In this limit the distribution function is that of the
isothermal sphere. 

It is unclear whether the degenerate limit, where $n_a \simeq \nu_a$,
is relevant for the cores of elliptical galaxies in the absence of
black holes. However, this limit is relevant in the presence of a
central black hole, as already pointed out by Stiavelli (1987).  In the
presence of a massive black hole of mass $M_{BH}$, one has that
$\Phi^{(ext)}(r_a) = -G M_{BH}/r_a$, having indicated with $r_a$ the
radial coordinate of macrocell $a$. This component dominates the
underlying gravitational potential at sufficiently small
radii. Consequently, for large portions of phase space $\exp{(\beta
E_a)}<<1$, and $n_a \simeq \nu_a$, i.e. the system becomes degenerate
at small radii. I present the resulting models in the next section.
It should be noted that incompleteness does not affect my conclusions,
since we are dealing with the nuclear galactic regions, where
relaxation is expected to be complete.

\section{Violent Relaxation in the Presence of a Massive Black Hole}

Let's rewrite for convenience the distribution function in continuous
representation by dropping the macrocell indices:
\begin{equation}
\label{feq}
f(v,r) = \frac{A}{1+\exp{(\beta E -\mu)}}
\end{equation}
with $A$ a normalization constant. At large
radii the non-degenerate limit applies, and the density and velocity
dispersion are those of the isothermal sphere:
\begin{equation}
\label{densityeq}
\rho = 4 \pi \int_0^{\infty} f(v,r) v^2 dv \propto r^{-2}
\end{equation}
and
\begin{equation}
\label{sigmaeq}
\sigma^2 = \frac{4 \pi}{ 3 \rho} \int_0^{\infty} f(v,r) v^4 dv =
\beta^{-1}.
\end{equation}

At radii sufficiently small so that the black hole dominates the
gravitational potential, one can estimate the integrals in
Eq. (\ref{densityeq}) and (\ref{sigmaeq}) with the saddle point method
to find $\rho \propto r^{-3/2}$ and $\sigma^2 \propto r^{-1}$.  These
slopes are equal to those obtained for models where the black hole
grows adiabatically in the core of an isothermal sphere. Black holes
grown adiabatically in non-isothermal models produce steeper cusps
(Quinlan, Hernquist and Sigurdsson 1995), which are therefore
distinguishable from those of the models presented here. In order to
carry out a more detailed comparison, I have computed a number of
self-consistent models from the distribution function of
Eq. (\ref{feq}) for a variety of degeneracy parameters $\mu$ and of
black hole masses.

In Figure 1 and 2 I show the comparison between the projected surface
brightness and velocity dispersion for {\it (i)} violent-relaxation,
degenerate models (solid lines), and for {\it (ii)} adiabatic black
hole growth models (squares).  The latter refer to an initial cored
isothermal sphere and have been computed by using software kindly made
available to me by G.D. Quinlan, and described in Quinlan et
al. (1995; for the model properties see also Young 1980; Lee and
Goodman 1989; Cipollina and Bertin 1994). I have assumed black hole to
galaxy core mass ratios of $M_{BH}/M_{core} = 0.03,0.1,0.3,0.5$.  In
the violent relaxation models, the value of the degeneracy parameter
$\mu$ depends on the initial conditions prior to the formation of each
galaxy, and therefore it is an additional free parameter. Numerical
experiments of dissipationless galaxy formation (without a central
black hole) indicate that the maximum final phase space density is
close to that of the initial conditions (e.g., Londrillo, Messina, and
Stiavelli 1991). This suggests that $\mu$ is of order unity. The
models of Figures 1 and 2 have been obtained for $\mu=0.5$. No best
fit to the adiabatic growth models has been attempted. If $\mu$
varies, the slope of the inner surface brightness profile is affected:
larger values of $\mu$ provide flatter slopes, and smaller values
provide steeper slopes. This ensures that the agreement between the
models presented here and the adiabatic models could be improved if
$\mu$ were treated as a free parameter. For the case of
$M_{BH}/M_{core}=0.3$, I show in Figure 3 the projected surface
brightness profiles for $\mu=0,0.2,0.5,0.7,1$.

The effect of degeneracy increases with increasing $\mu$.
For values of $\mu << 0$ the models become increasingly similar to the
isothermal sphere without a black hole but show a bi-modal density
profile, with an overluminous cusp, when a black hole potential is
included. Models with $\mu >> 1$ are degenerate even in the absence of
a black hole.  

Both the adiabatic models and the maximum entropy models ignore two
physical effects: {\it i)} the fact that the black hole may be not at
rest in the center and, {\it ii)} stellar disruption by the black
hole. For the models presented here, the former should probably not
affect significantly the end result which does not depend on the
detailed fluctuations of the gravitational potential. The latter would
introduce a tangential anisotropy in the orbital distribution close to
the black hole and alter the cusp profile at very small radii. A
precise numerical evaluation of these effects goes beyond the scope of
this paper.

The main conclusion that one can draw from this study is that there is
a fairly good agreement between adiabatic growth models and
degenerate, violent relaxation models, for the {\it same} values of
black hole to galaxy core mass ratio.

\section{Conclusions}

Based on statistical mechanical arguments, I have derived the
distribution function which results from violent relaxation in the
presence of a pre-existing massive black hole, i.e. in the case where
the formation of the black hole acts as a seed for the formation of
the galaxy. The corresponding stellar dynamical models are
qualitatively similar to models of adiabatic growth of a central black
hole from an isothermal core, i.e. the physically opposite scenario,
where the formation of the galaxy precedes that of the central black
hole. For appropriate values of the degeneracy parameter $\mu$, which
is arbitrary in the description that I have presented, the two
scenarios become essentially indistinguishable. It is unlikely that
the models could be distinguished on the basis of high spatial
resolution line profiles since the central tangential anisotropy of
adiabatic models is small and, in any case, stellar distruption could
produce a similar anisotropy also in the maximum entropy models.  This
implies that on the basis of fits to the observed black hole cusps in
galaxies, we are unable to infer the formation mechanism of the black
hole and of its cusp.

\section{Acknowledgements}

I thank Giuseppe Bertin, Marcella Carollo and Roeland van der Marel
for useful discussions, and Gerald Quinlan for kindly providing the
code to compute adiabatic growth models. The anonymous referee provided
useful comments. This work has been partially supported by MURST of
Italy and by the Italian Space Agency (ASI).

\newpage
\section{Figure Captions}

{\bf Figure 1:} Comparison of the projected surface brightness between
my models ($\mu=0.5$, solid lines), and adiabatic black hole growth
models (squares).  Plotted are curves for black hole to galaxy core
mass ratios $M_{BH}/M_{core} = 0.03,0.1,0.3,0.5$. There is a fairly
good agreement between the two families of models for large
$M_{BH}/M_{core}$ values. The radius $R$ is measured in units of the
core radius.

{\bf Figure 2:} Comparison of the projected projected velocity dispersion
between my models ($\mu=0.5$, solid lines), and models of adiabatic
black hole growth (squares).  The black hole to galaxy core mass
ratios are $M_{BH}/M_{core} = 0.03,0.1,0.3,0.5$. The radius $R$ is in
units of the core radius. The model without a black hole, not shown in
the figure, has a velocity dispersion profile characterized by the
presence of a small step, which occurs at the radius where degeneracy
starts affecting the distribution function. A residual of this effect
is seen for small values of $M_{BH}/M_{core}$. It disappears for
$M_{BH}/M_{core} \simeq 0.3$.

{\bf Figure 3:} Projected surface brightness profiles for degenerate,
violent relaxation models with $M_{BH}/M_{core} = 0.3$, and degeneracy
parameter $\mu=0$ (dot-dashed), $0.2$ (long dashed), $0.5$ (solid),
$0.7$ (short dashed), and $1$ ( dotted).

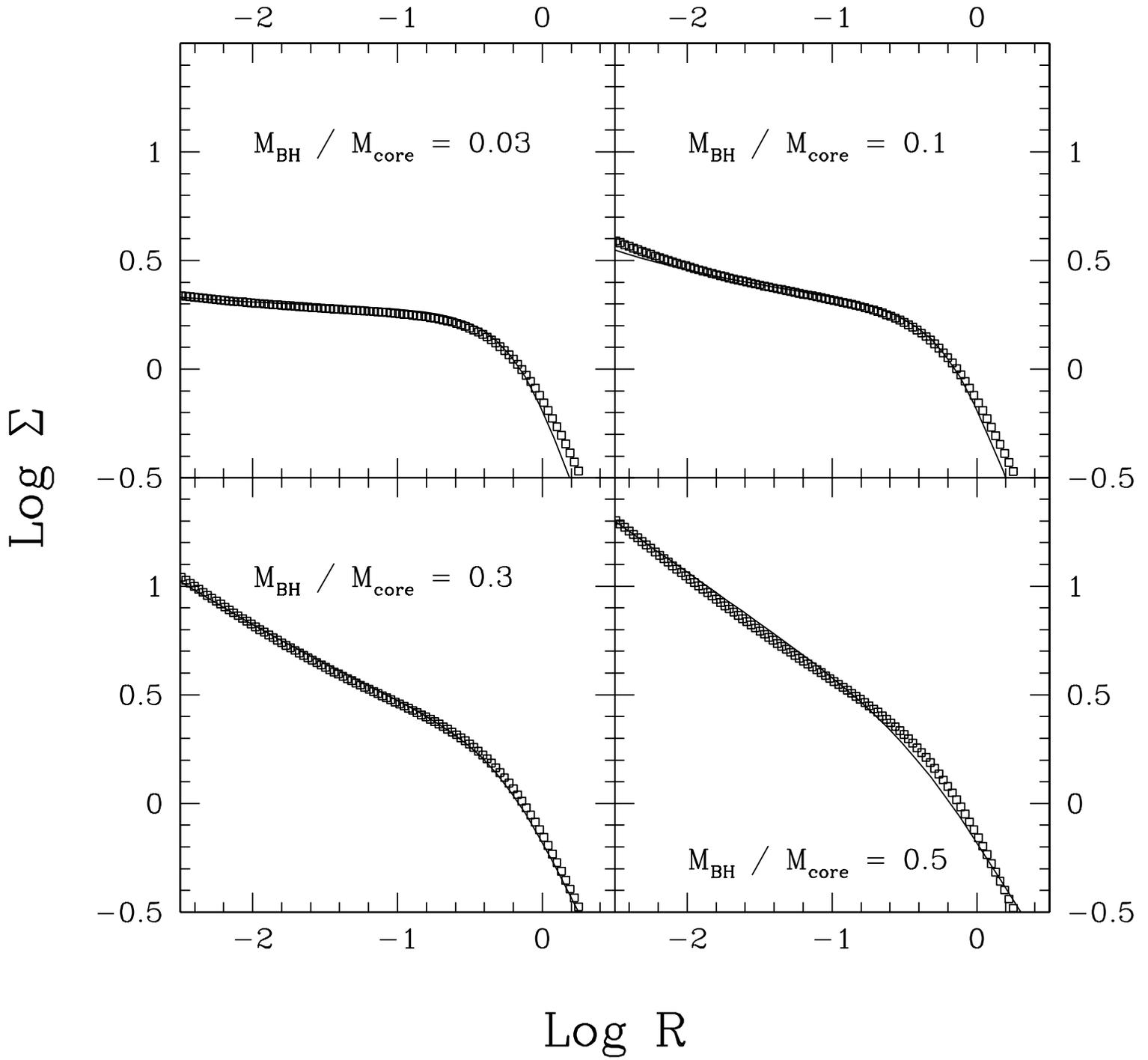
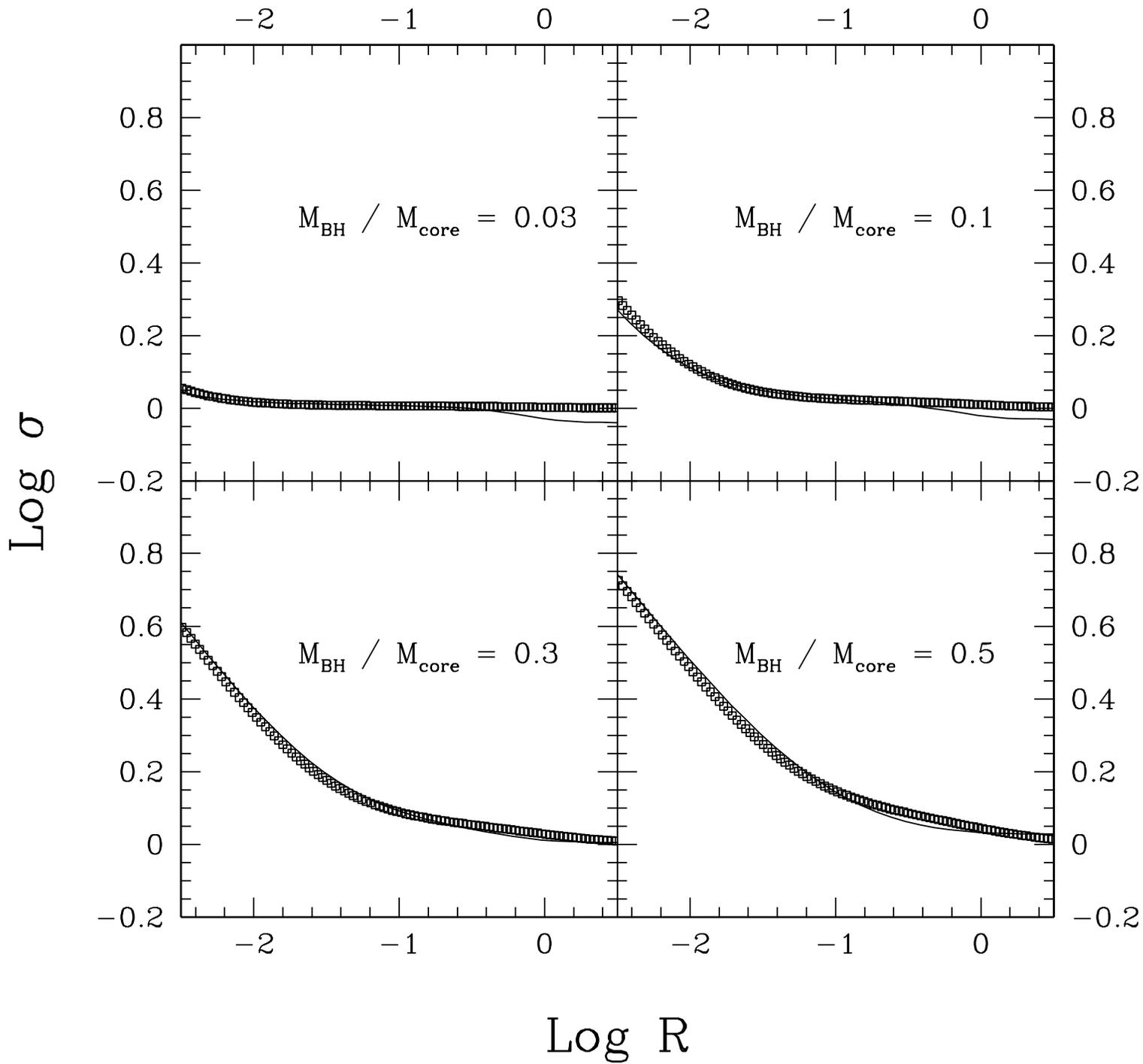
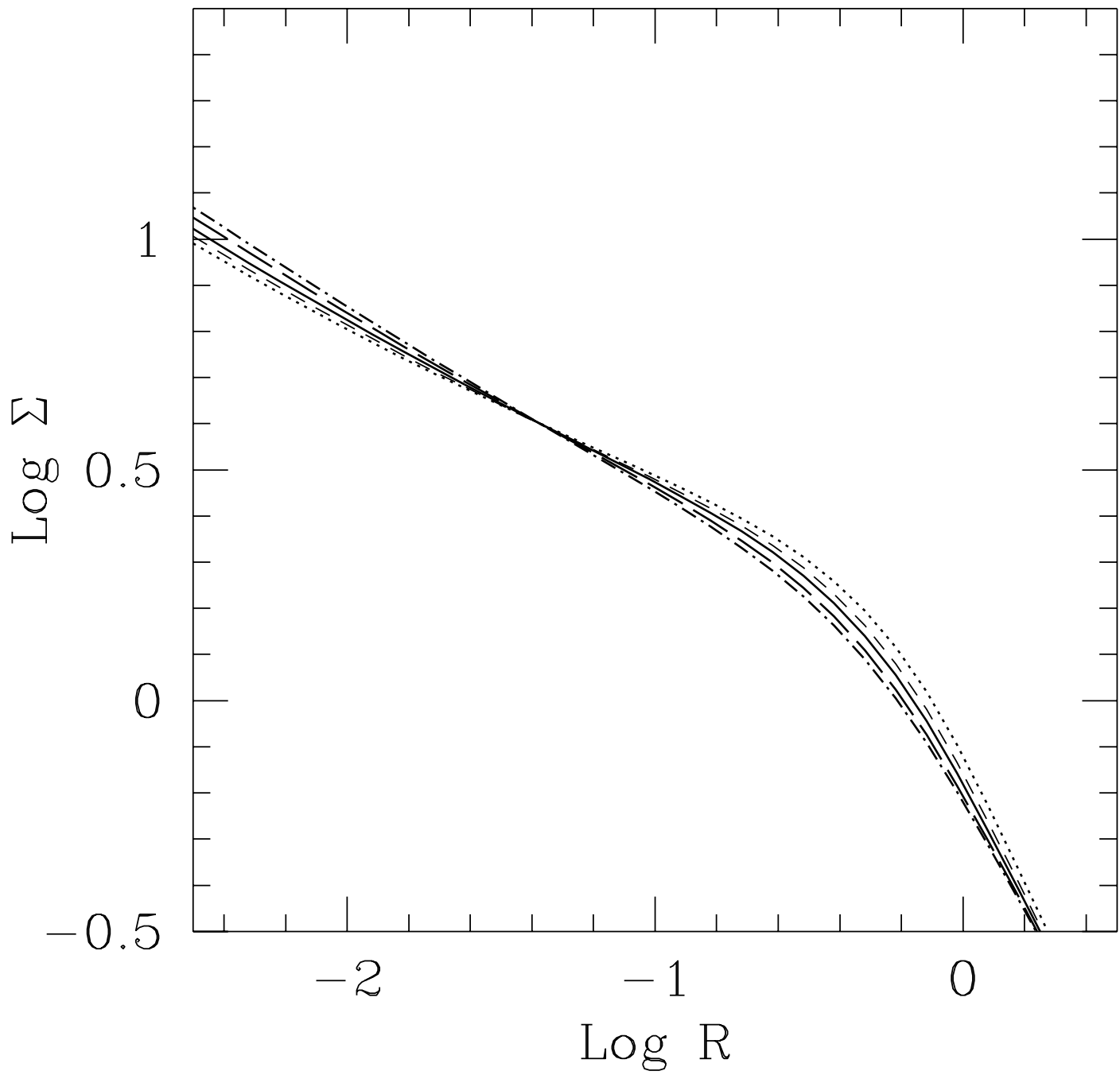
\end{document}